\documentclass[aps,prl,twocolumn,amsmath,amssymb,floatfix, showpacs]{revtex4}
\usepackage{graphicx,bm,ulem,dcolumn}
\usepackage[usenames,dvipsnames]{color}

\begin{document}
\title{Turbulence comes in bursts in stably stratified flows}

\author{C. Rorai$^{1,2}$,  P.D. Mininni$^{3}$ and A. Pouquet$^{4,5}$}
\affiliation{
$^1$Nordita, Roslagstullsbacken 23, 106 91 Stockholm, Sweden; \\
$^2$ICTP, Strada Costiera 11, 34151 Trieste, Italy;  \\
$^3$Departamento de F\'\i sica, Facultad de Ciencias Exactas y Naturales, Universidad de Buenos Aires; \\ 
$^4$National Center for Atmospheric Research, P.O. Box 3000, Boulder CO 80307, USA; \\
$^5$Department of Applied Mathematics, CU, Boulder, CO, 80309-256 USA.
}
\begin{abstract}
There is a clear distinction between simple laminar and complex turbulent fluids. But in some cases, as for the nocturnal planetary boundary layer, a stable and well-ordered flow can develop intense and sporadic bursts of turbulent activity which disappear slowly in time. This phenomenon is ill-understood and poorly modeled; and yet, it is central to our understanding of weather and climate dynamics. We present here a simple model which shows that in stably stratified turbulence, the stronger bursts can occur when the flow is expected to be more stable. The bursts are generated by a rapid non-linear amplification of energy stored in waves, and are associated with energetic interchanges between vertical velocity and temperature (or density) fluctuations. Direct numerical simulations on grids of $2048^3$ points confirm this somewhat paradoxical result of measurably stronger events for more stable flows, displayed not only in the temperature and vertical velocity derivatives, but also in the amplitude of the fields
themselves.
\end{abstract}


\maketitle

Large fluctuations are common in physical systems with long-range correlations, and have been found to be linked to so-called ``1/f'' noise \cite{antal_01}. They take the form of sporadic and localized events, as observed in many instances in critical phenomena and in turbulent flows, and  are diagnosed through non-Gaussian Probability Distribution Functions (PDFs) \cite{bramwell_00}. In turbulence, strong events occur in field gradients, with the velocity itself being nearly Gaussian. There are however exceptions to this last rule for shear flows \cite{pumir_96, marino_12}, quantum fluids \cite{paoletti_08,baggaley_11}, and subtropical current systems \cite{capet_08}. Extreme events associated with random plumes have also been diagnosed in the atmospheric convective boundary layer  \cite{sullivan_11}, or when linked with coherent structures, e.g., storm tracks. 

The occurrence of intermittent strong activity is a signature of fully developed turbulence and is therefore more surprising in stable flows. Intermittency makes the  stable nocturnal planetary boundary layer (PBL) highly unpredictable: as night sets in, this layer between the atmosphere and land or sea stabilizes due to the radiative cooling of the land and ocean masses. It is still unclear how stable the nocturnal PBL becomes. Three regimes have been observed \cite{jielun_12}: very stable, weakly stable with turbulent motions persisting and competing with internal gravity waves, and transitory. Even in the very stable case, the PBL is subject to intense sporadic bursts of turbulence which die out after many wave periods  \cite{finnigan_99, jielun_12}. 

Numerical simulations play an increasing role in the understanding of these complex processes, and in quantifying the dual problem of the increased stability \cite{jonker_12a} and the spontaneous generation of bursts. However, modeling of the PBL in weather and climate codes is often inadequate, resulting, for example, in an inaccurate evaluation of the extension of the ice sheet, as is the case over Greenland \cite{drue_07}, and in a faulty estimate of the overall energy balance in long-term climate systems, since it affects for example mixing, frost occurrence, aerosol dispersion, and air quality \cite{stevens_09}.
 
{\it The model.} We start from the Boussinesq equations, which describe a stably stratified flow with gravity in the vertical direction. For the velocity ${\bf u}=(u,v,w)$ and potential temperature fluctuations $\theta$, the equations are
\begin{eqnarray} 
\partial_t {\mathbf u} +{\mathbf u} \cdot \nabla {\mathbf u}  &=&  -\nabla P - N \theta\  e_z + \nu \Delta {\mathbf u} \ , \label{eq:vel} \\
\partial _t \theta\  +{\mathbf u} \cdot \nabla \theta\  &=& N w + \kappa \Delta \theta\  \   \ \ \ , \ \  \ \nabla \cdot {\bf u} = 0 \ , \label{eq:temp}
\end{eqnarray} 
where $P$ is the pressure, and $\kappa=\nu$ the diffusivity. The square Brunt-V\"ais\"al\"a frequency is given by $N^2=-(g/\theta ) (d\bar \theta /dz)$, where $d\bar \theta /dz$ is the imposed background stratification, assumed to be linear, and $g$ is the gravity.

Estimating the pressure forces, which for an incompressible fluid are highly non-local, is difficult since one has to consider the coupling between vorticity and shear. A simple model of such behavior was developed in \cite{vieille}. This model, sometimes called ``restricted Euler dynamics'', has proven useful in analyzing the development and the statistical and geometrical properties of intermittent structures in a variety of turbulent flows \cite{meneveau_11}.
 
For simplicity, in the absence of stratification one can consider only vertical velocity differences $\delta w$ in the vertical velocity $w$ at scale $\ell$, defined as
$ 
\delta w (\ell) = \left< w ({\bf x}+\ell \hat{z})-w({\bf x}) \right> \approx \ell \partial_z w .
$ 
Taking the spatial derivative of Eq.~(\ref{eq:vel}) in the one dimensional (1D) case, with $\theta=0$, and neglecting pressure and viscous forces
yields:
$$
\partial_t (\partial_z w) + w \partial_z (\partial_z w) = d_t (\partial_z w) = -(\partial_z w)^2.
$$
Then, for the velocity differences $ d_t \delta w =- \delta w^2/\ell$. This equation immediately shows the temporal enhancement of negative values of $\delta w$, as observed for example for isotropic turbulent fluids for which the skewness of velocity gradients is negative and of order unity.

When the flow is stably stratified, gravity acts as a restitutive force allowing for oscillatory solutions (internal gravity waves). Non-linear coupling tends to transfer energy towards modes with vertical spatial dependence, resulting in the creation of horizontal layers in the fluid, and further justifying the reduction to a 1D system. Under the same hypothesis, for $\delta \theta \approx \ell \partial_z \theta$, and from Eqs.~(\ref{eq:vel}) and (\ref{eq:temp}), we obtain
\begin{equation}
\frac{d \delta w}{dt} = - \frac{\delta w ^2}{\ell} - N \delta \theta , \ \ \frac{d \delta \theta}{dt} = - \frac{\delta w \delta \theta}{\ell} + N \delta w  \ .
\label{model_eq}  \end{equation}
%%%%%%%%%%%%%%%%%%%% 
\begin{figure}
\centering
\resizebox{8.9cm}{!}{\includegraphics{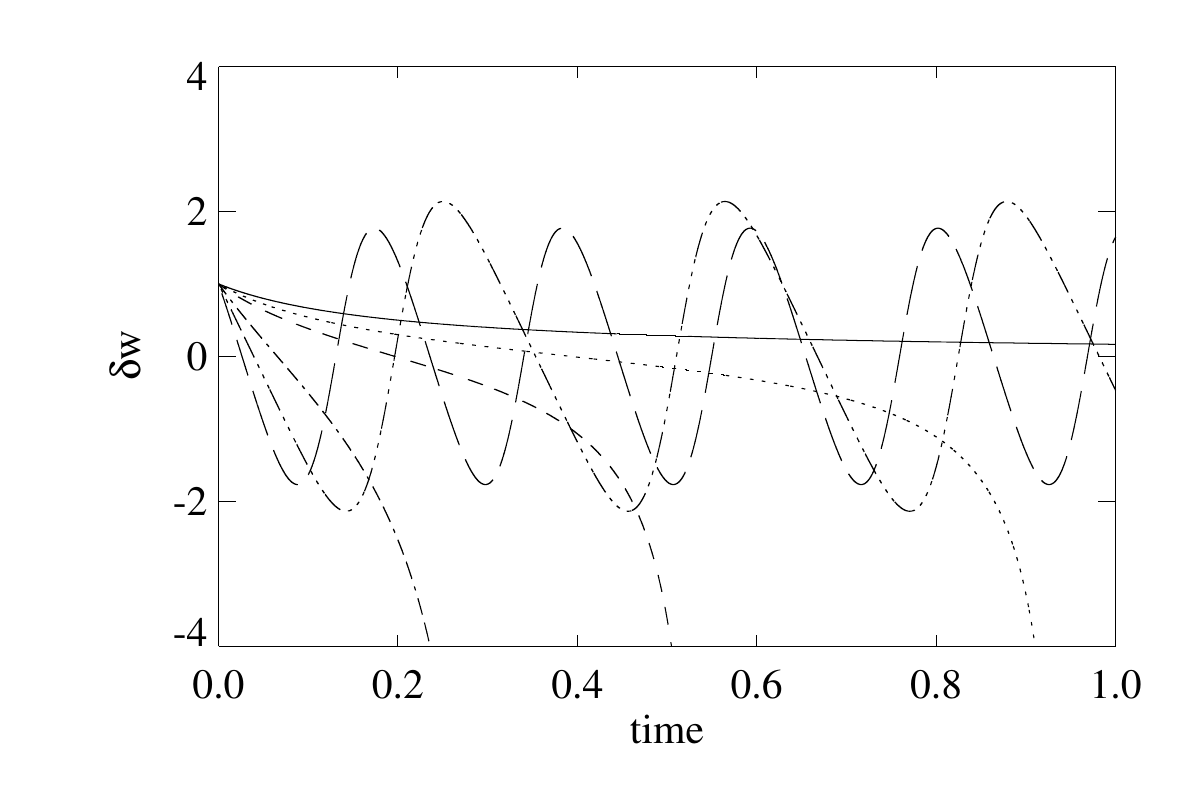}}
\caption{\label{model_fig}
Evolution in time of vertical velocity variations $\delta w$ in the model of Eq.~(\ref{model_eq}) for $\ell=0.2$ and $N=0$ (no stratification, solid line), and for $N=2$ (dotted), 4 (dashed), 12 (dash-dotted), 20 (dash-triple-dotted), and 30 (long dashed line). Note the faster evolution towards negative and strong vertical gradients at intermediate values of $N$, before oscillatory behavior takes over for large enough $N$.
} \end{figure} 
%%%%%%%%%%%%%%%%%%%%  

These equations can be considered as a crude 1D model of a stratified flow. We can define the dimensionless Froude number $Fr=U/(NL)$ (with $U$ and $L$  characteristic velocity and length); it quantifies the ratio between nonlinear and linear effects. System (\ref{model_eq}) has only one fixed point ($\delta w = \delta \theta=0$). For weak stratification, one recovers the Euler behavior of strong negative gradients, and in the opposite case ($N \gg 1$), the model has oscillatory solutions in the vertical velocity and temperature fluctuations (see Fig.~\ref{model_fig}). 

\begin{figure}
\centering
\resizebox{8.3cm}{!}{\includegraphics{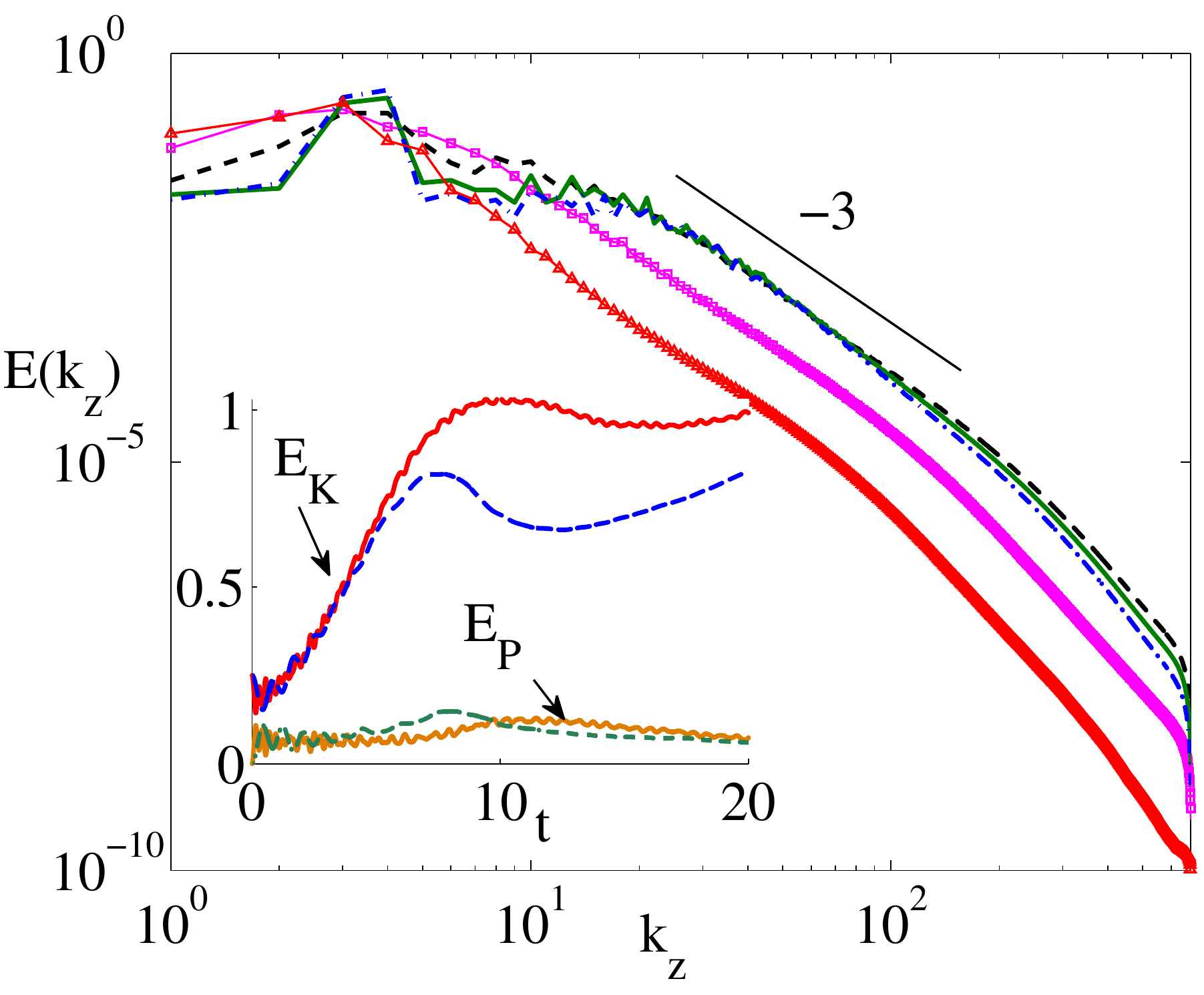}}
\resizebox{8.3cm}{!}{\includegraphics{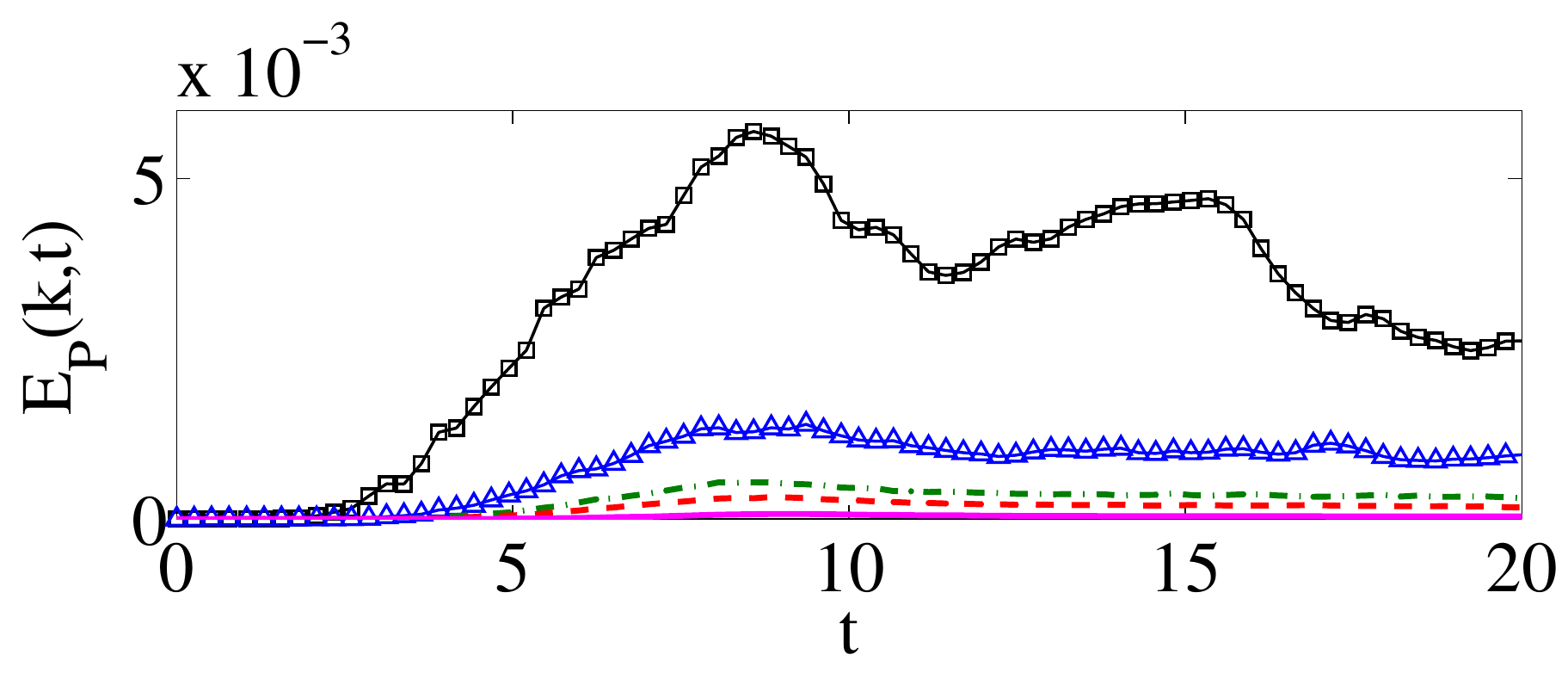}}
\resizebox{8.3cm}{!}{\includegraphics{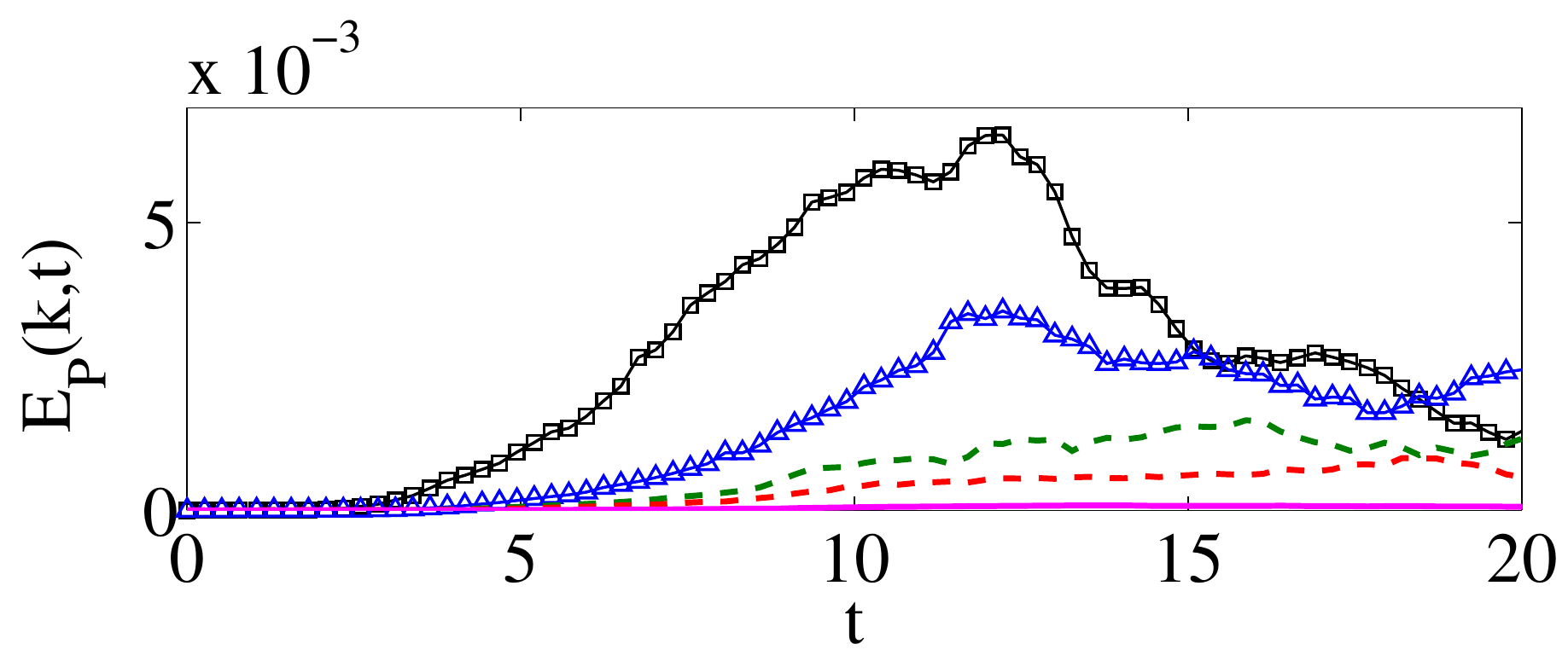}}
\caption{\label{spectra}
({\it Color online.)} 
{\underline{{\it Top:}}}
Parallel energy spectrum for the run with $N=12$. Times correspond to $t=4.94$ (triangles, red), $t=7.54$ (squares, magenta), and $t>12$ for the remaining curves, when small scales have reached a turbulent steady state. A $\sim k_z^{-3}$ scaling is shown as a reference. The inset shows the time evolution of the potential energy $E_P$ and of the kinetic energy $E_K$ in runs with $N=4$ (solid) and $N=12$ (dashed). Note the oscillations associated with internal gravity waves for $N=12$.
{\underline{{\it Middle:}}}
Time evolution of the potential energy for $k=$10 (squares, black), 20 (triangles, blue), 30 (dash-dotted, green), 40 (dashed, red), and 80 (solid, magenta), in the run with $N=4$.
{\underline{{\it Bottom:}}}
 Same for the run with $N=12$. Note the larger fluctuations and bursts for large $k$ in this run reflected by the values of the standard deviation:
 $\sigma_{N=4}( k=10) =1.0 \times10^{-3}$, $\sigma_{N=12}(k=10) = 8.4 \times 10^{-4}$, $\sigma_{N=4}(k=40) =4.2 \times 10^{-5}$, $\sigma_{N=12}(k=40) = 1.1\times 10^{-4}$.
} \end{figure}

The terms governing both (non-linear and linear)  behaviors  become comparable when $\delta w \sim \delta \theta \sim N\ell$. When this is satisfied in a range of scales, it corresponds to the balanced energy spectrum $E(k_z)\sim \delta w^2/k_z \sim N^2 k_z^{-3}$ which has been predicted and observed in many instances in the atmosphere and the oceans (see, e.g., \cite{polzin} and Fig.~\ref{spectra}).

%%%%%%%%%%%%%%%%%%%% 
%%%%%%%%%%%%%%%%%%%%  

%%%%%%%%%%%%%%%%%%%% 
\begin{figure}
\centering
\resizebox{8.1cm}{!}{\includegraphics{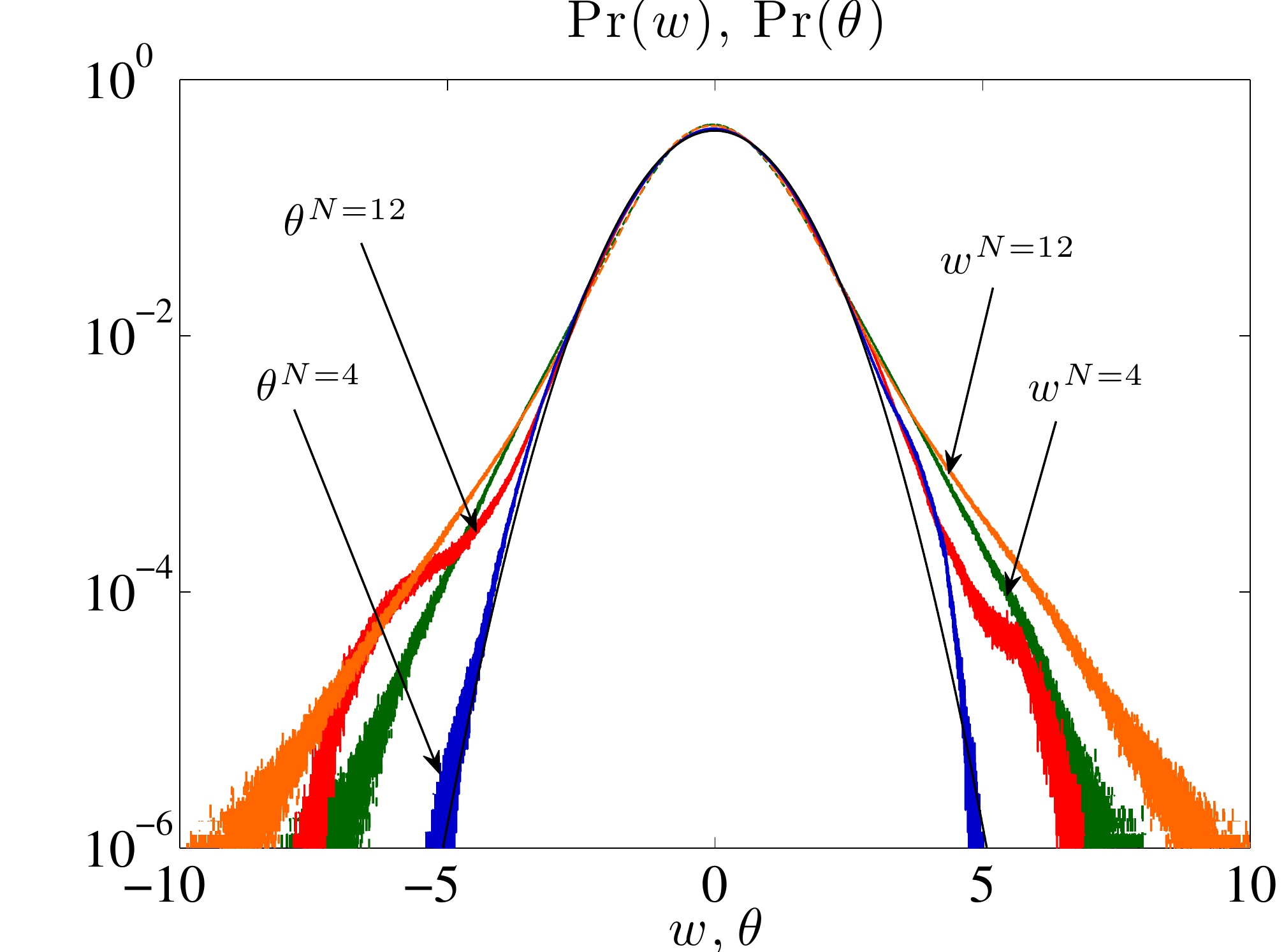}}
\resizebox{8.3cm}{!}{\includegraphics{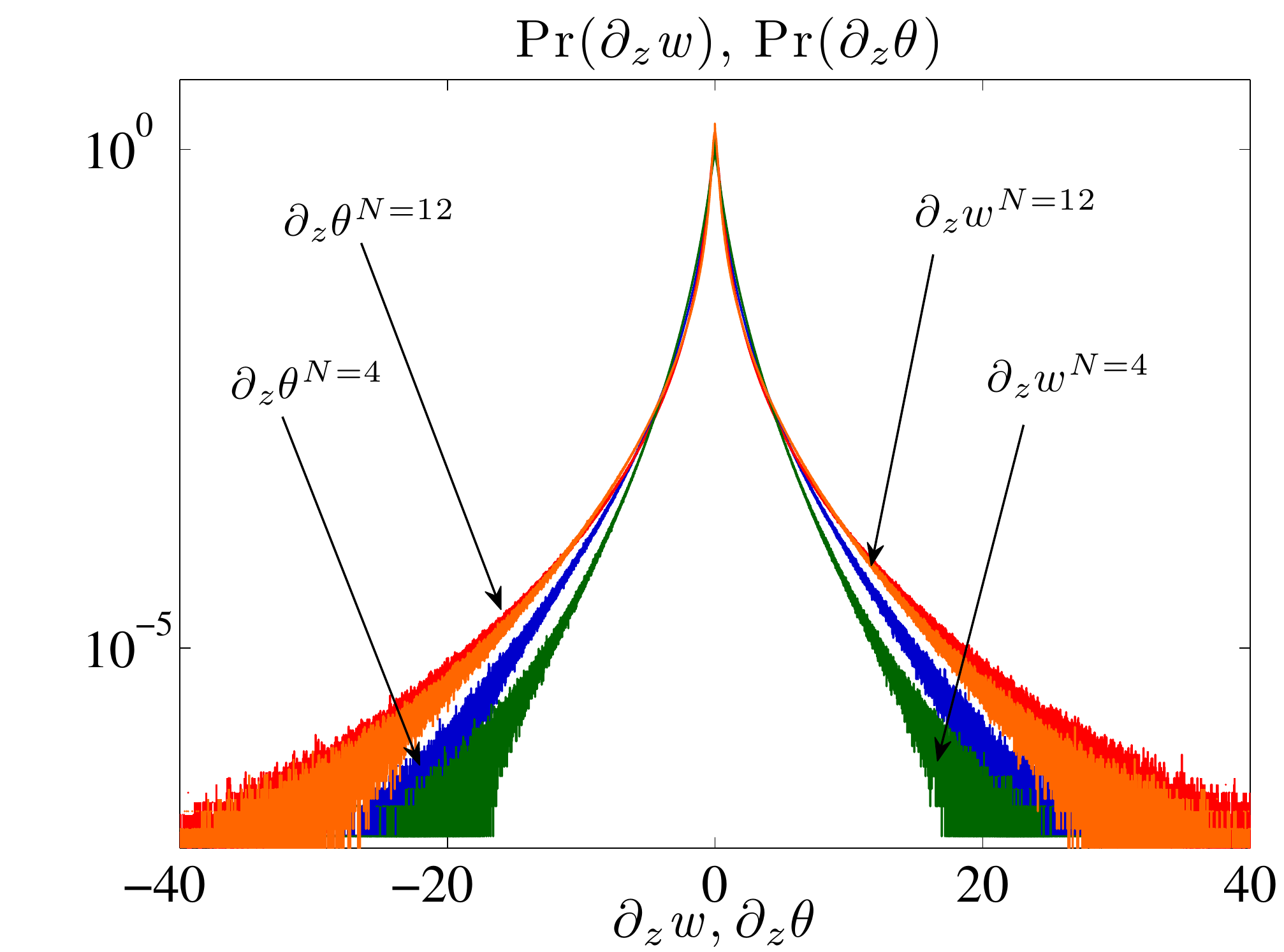}}
\caption{\label{pdfs}
{\it Above:} ({\it Color online.)}
Normalized histograms (in semi-log coordinates) evaluated shortly after the peak of dissipation, for the temperature fluctuations $\theta$ and for the vertical component of the velocity, $w$,  for high resolution simulations of a stratified flow with Froude number $Fr\approx 0.1$ ($N=4$) and $Fr \approx 0.03$ ($N=12$).
A normal distribution is shown (inner black curve) as a reference. 
{\it Below:} PDFs of vertical derivatives for the same quantities. In all cases, the more strongly stratified flow with $N=12$ has larger probability of developing extreme events, as illustrated by the wider wings in the PDFs. For the fields themselves, the velocity is more intermittent than the temperature, and the converse is true for their vertical derivatives.} \end{figure} 
%%%%%%%%%%%%%%%%%%%%  

In Fig.~\ref{model_fig}, an interesting evolution is observed in this intermediate regime: for $N=2$, 4 and 12, and for initial $\delta w$ and $\delta \theta >0$, $\delta w$ becomes negative (and diverges) unlike the case $N=0$, and faster for larger values of $N$. In other words, waves are amplified by the nonlinear term, resulting in a catastrophic behavior. The run-away occurs as $N$ increases and before oscillations take over, in Fig.~\ref{model_fig} for $ N > 12$. The large negative values of $\delta w $ can be interpreted as the signature of strong intermittent bursts. Note that for larger values of $N$, although the solutions become oscillatory, they still display skewness (i.e., they have a preference towards more negative values of $\delta w$). If the initial conditions are negative ($\delta w$, $\delta \theta <0$), the divergence is delayed by increasing stratification.

%%%%%%%%%%%%%%%%%%%% 
\begin{figure} \centering
\resizebox{8.5cm}{!}{\includegraphics{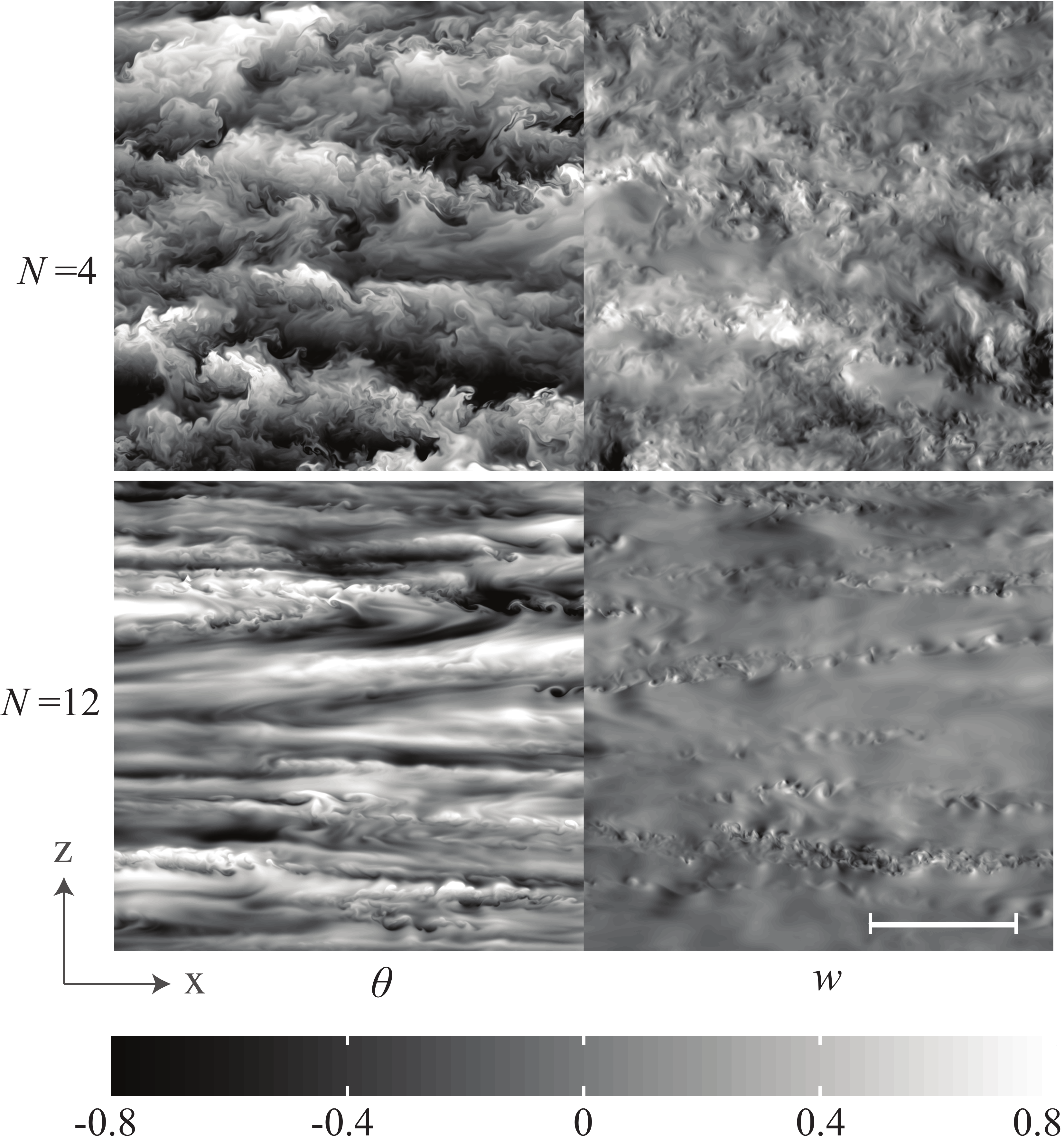}}
\resizebox{8.5cm}{!}{\includegraphics{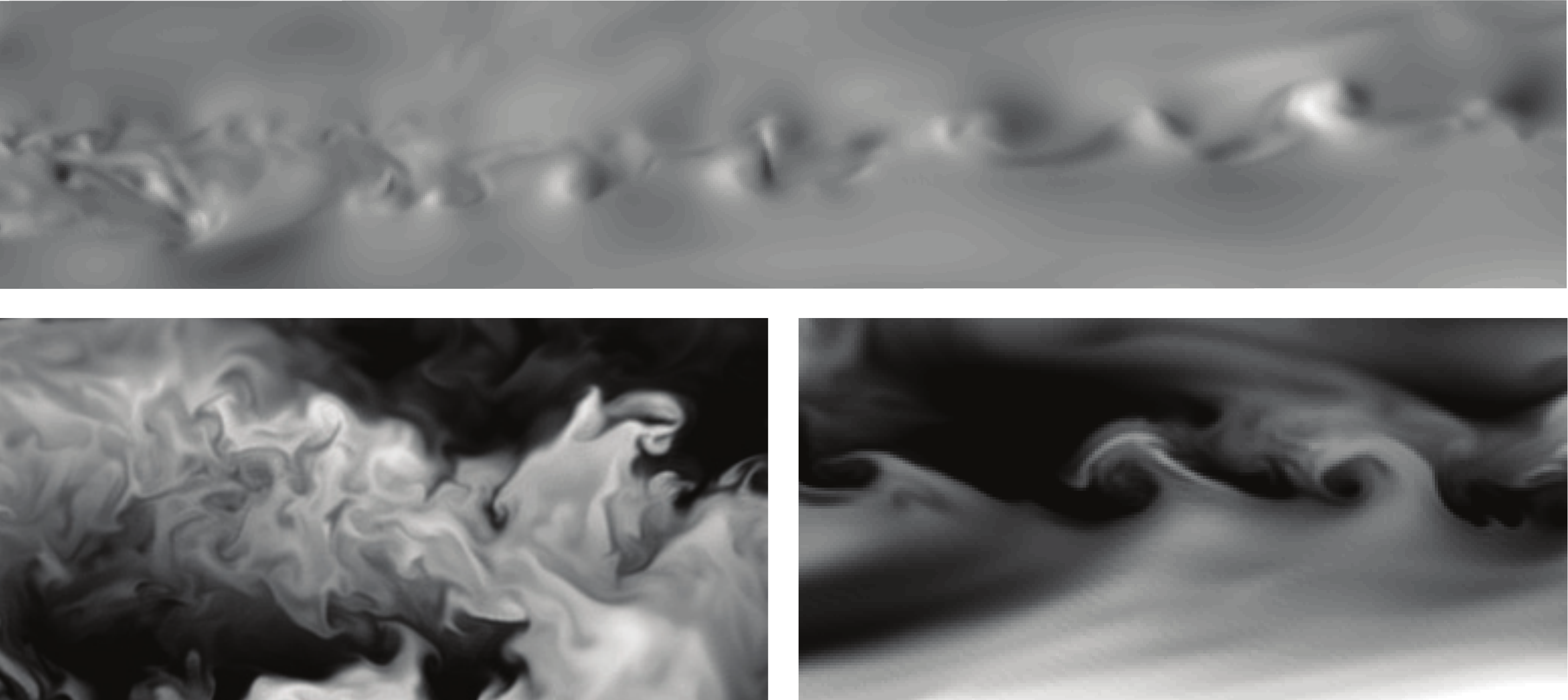}}
\caption{\label{slices}
{\it Above:} Two-dimensional cuts in a $[0, 2\pi]^3$ box in the $x,z$ directions for the flow with $Fr\approx 0.1$ ($N=4$, top rows) and $Fr\approx 0.03$ ($N=12$, bottom rows). The white segment is of unit length. The first column corresponds to the temperature, and the second to the vertical velocity. Note the strata in the temperature (the more strata, the higher the value of $N$). For $Fr\approx 0.03$, sporadic overturning in the vertical cut and weaker eddies in the horizontal plane are clearly visible. The flow with stronger stratification is less complex, but extreme values of the fields and their gradients are higher, leading to the development of turbulent bursts and localized mixing.
{\it Below:} Details of some extreme events: Kelvin-Helmholtz (KH) instability in the velocity field for $Fr\approx 0.03$ ($N=12$, top),  eddies in the temperature field for $Fr\approx 0.1$ ($N=4$, bottom left), KH instability in the temperature field for $N=12$ (bottom right).
} \end{figure} 
%%%%%%%%%%%%%%%%%%%%

The coupling of this evolution to that of the horizontal velocity damps the run-away evolution of $\delta w$ (because of incompressibility) but strong gradients still form (see \cite{li_meneveau_06} for a similar model for non-stratified flows with a passive scalar, and \cite{li_10} for rotating flows which  display less extreme events than isotropic  homogeneous turbulence). Our model can also be extended to consider the effects of shear in the flow, and results in a stronger amplification of velocity variations.

This run-away toward strong gradients can be interpreted somewhat differently: for a given level of stratification, there exists a scale $\ell$ at which strong negative tails in the velocity fluctuations will occur. For $N$ not too large, extreme events can develop even at large scale, and be visible in the PDFs of the fields themselves. This behavior could be linked to the phenomenon of non-linear amplification of waves observed in the nocturnal PBL \cite{finnigan_99, jielun_12} and in flows with internal shear and density fluctuations, as in the solar wind \cite{marino_12}.

{\it Numerical simulations.} In order to ascertain the value of the model written above, we now examine the dynamics of stratified turbulence using direct numerical simulations (DNS) at high resolution. To sustain the flow, we add a three-dimensional large-scale random isotropic forcing to the r.h.s.~of Eq.~(\ref{eq:vel}). Equations (\ref{eq:vel}) and (\ref{eq:temp}) are solved on grids of $2048^3$ points with the pseudo-spectral Geophysical High-Order Suite for Turbulence code, which is parallelized with hybrid MPI/OpenMP programming, and has been tested on over 98000 compute cores  \cite{hybrid2011}. As the amplification in the simplified model above can happen within the fluid, we consider for simplicity tri-periodic boundary conditions within a $[0,2\pi]^3$ box, with a $2^{nd}$--order explicit Runge-Kutta temporal scheme, and with de-aliasing using a standard 2/3 rule.

Two simulations were conducted for over 20 turnover times, $\tau_{NL}=U/L$. In both simulations, the flow was started from random Gaussian isotropic initial conditions for the velocity, and with $\theta = 0$. The viscosity is such that the Reynolds number (the ratio of nonlinear to viscous effects) is $Re \approx 2.5 \times 10^4$ for both runs. The simulations have either $Fr \approx 0.1$ ($N=4$), or  $Fr \approx 0.03$ ($N=12$). Data for the statistical analysis is extracted soon after the peak of dissipation is reached, but similar results are obtained at different times after and before the peak.

Fig. \ref{spectra} shows the time evolution of the energy spectrum when $N=12$, which is compatible with $\sim k_z^{-3}$ scaling. In Fig.~\ref{spectra} we also show the time evolution of the kinetic and potential energies in both runs; the run with stronger stratification shows oscillations associated with internal gravity waves.

We examine now the occurrence of extreme events in these high resolution runs. As stratification increases, the flow is expected to become more stable and predictable, developing weaker events in the velocity and temperature. However, the opposite is observed. In Fig.~\ref{spectra} we 
also
show the time evolution of the potential energy at different wavenumbers. At the smallest scales (small $\ell \sim 1/k$, i.e., larger wavenumbers), the time series of the run with $N=12$ is more bursty than the run with $N=4$. By measuring the standard deviation of the time series, we verified that the time series at large scales in the run with $N=4$ have larger fluctuations than for $N=12$ (i.e., stratification smooths the evolution for sufficiently large $\ell$), while at small scales the opposite happens, in agreement with the qualitative behavior observed in the model. In particular, note that at $k=30$ and 40 the potential energy as a function of time is almost constant after $t\approx 10$ in the run with $N=4$, while it shows bursts and fluctuations in the run with $N=12$.

A better quantification of the strength of these events can be obtained from spatial information. In Fig.~\ref{pdfs} we show the PDFs for velocity and temperature fluctuations and their vertical gradients. We observe that: (i) for a given field, the more stratified case is more bursty, as illustrated by the heavy tails of the histograms which indicate a larger probability of the fields taking extreme values; (ii) the velocity is intermittent, more so than the temperature; conversely, (iii) the spatial derivative of the temperature takes larger extreme values than the derivative of the vertical velocity. Although non-Gaussian tails have been reported in PDFs of the field gradients \cite{brethouwer_07}, note that here the PDFs of the fields themselves are non-Gaussian, with the pointwise temperature and velocity taking extreme values.

Fig. \ref{slices} displays vertical slices of temperature and vertical velocity, for the same two flows. As expected, in the less stratified flow, turbulence in the form of multiple eddies, is ubiquitous. The more stratified flow seems more ordered, but strong small-scale Kelvin-Helmholtz type instabilities develop and dominate the small-scale dynamics as seen in several locations (see Fig. \ref{slices}); in this flow, turbulence comes in localized bursts with strong values of the fields and their gradients.

Note also the layered structures in the vertical, specially in the temperature. The number of layers depends on the stratification, with vertical correlation lengths of $\approx 1/8$ of the box in the flow with $Fr \approx 0.1$ ($N=4$), and $\approx 1/26$ of the box for $Fr \approx 0.03$ ($N=12$). If these values are used for $\ell$ in the model (\ref{model_eq}), together with the corresponding values of $N$, a faster growth of the fields is obtained for the linearly more stable flow, in good agreement with the DNS, and as indicated by observations.

{\it Discussion.} Our 
model and
computations point to stably stratified flows spontaneously developing long-lasting bursts which are stronger for stronger stratification. We can infer from our model that propagating gravity waves are non-linearly amplified, resulting in their breaking and the generation of turbulence when the linear and nonlinear effects are balanced. This is reminiscent of parametric instabilities, the forcing being provided by a spectrum of nonlinear eddies. Note that for  $E(k_z)\sim N^2 k_z^{-3}$, the eddy turn-over time is proportional to $1/N$, validating this balance across all scales \cite{billant_01, lindborg2006}. 

From the model and the simulations, we can conclude that the intermittent bursts are associated with the direct coupling between  vertical velocity and  temperature fluctuations. Our model thus allows for a simple explanation of the intriguing observation of strong intermittency in the stably stratified nocturnal planetary boundary layer.

It is known that wave turbulence can lead to intermittency, such as with rogue waves in the ocean; anomalous concentrations of particles in large-scale waves \cite{falko_05b}, and enhancement of non-Gaussian initial conditions \cite{lvov_04} have been observed as well. In a turbulent flow, intermittency has been related to a global correlation between interacting scales, and as such may be indicative of a system close to criticality \cite{portelli_03}. Similarly, non-locality of interactions between small-scale and large-scale eddies is advocated in \cite{fritts_09a} as directly related to the bursting phenomenon of the nocturnal PBL. Our model indicates a different origin for the strong localized events, associated with a positive feedback in the vertical between nonlinear steepening and wave motions, and is consistent with the simulations that indicate that a more stably stratified flow has stronger bursts in a plage of parameters. This study may thus lead to more useful parametrizations of stably stratified flows in weather and climate models, by formulating a stochastic eddy-noise \cite{palmer_12} which explicitly incorporates the non-linear coupling described herein (for a quasi-normal closure, see \cite{gryanik_05}).

{\it This work was supported by NSF/CMG 1025183. C. Rorai acknowledges support from two RSVP/CISL grants. Computer time on Yellowstone through ASD was provided by NCAR under sponsorship of NSF. AP gratefully acknowledges useful discussions with the WINABL group and in particular with J.~Finnigan and P.~Sullivan.}

\bibliography{ms.bbl}

\begin{thebibliography}{26}
\expandafter\ifx\csname natexlab\endcsname\relax\def\natexlab#1{#1}\fi
\expandafter\ifx\csname bibnamefont\endcsname\relax
  \def\bibnamefont#1{#1}\fi
\expandafter\ifx\csname bibfnamefont\endcsname\relax
  \def\bibfnamefont#1{#1}\fi
\expandafter\ifx\csname citenamefont\endcsname\relax
  \def\citenamefont#1{#1}\fi
\expandafter\ifx\csname url\endcsname\relax
  \def\url#1{\texttt{#1}}\fi
\expandafter\ifx\csname urlprefix\endcsname\relax\def\urlprefix{URL }\fi
\providecommand{\bibinfo}[2]{#2}
\providecommand{\eprint}[2][]{\url{#2}}

\bibitem[{\citenamefont{Antal et~al.}(2001)\citenamefont{Antal, Droz,
  Gy{\"o}rgyi, and R{\`a}cz}}]{antal_01}
\bibinfo{author}{\bibfnamefont{T.}~\bibnamefont{Antal}},
  \bibinfo{author}{\bibfnamefont{M.}~\bibnamefont{Droz}},
  \bibinfo{author}{\bibfnamefont{G.}~\bibnamefont{Gy{\"o}rgyi}},
  \bibnamefont{and} \bibinfo{author}{\bibfnamefont{Z.}~\bibnamefont{R{\`a}cz}},
  \bibinfo{journal}{Phys. Rev. Lett.} \textbf{\bibinfo{volume}{87}},
  \bibinfo{pages}{240601} (\bibinfo{year}{2001}).

\bibitem[{\citenamefont{Bramwell et~al.}(2000)\citenamefont{Bramwell,
  Christensen, Fortin, Holdsworth, Jensen, Lise, López, Nicodemi, Pinton, and
  Sellitto}}]{bramwell_00}
\bibinfo{author}{\bibfnamefont{S.~T.} \bibnamefont{Bramwell}},
  \bibinfo{author}{\bibfnamefont{K.}~\bibnamefont{Christensen}},
  \bibinfo{author}{\bibfnamefont{J.-Y.} \bibnamefont{Fortin}},
  \bibinfo{author}{\bibfnamefont{P.~C.~W.} \bibnamefont{Holdsworth}},
  \bibinfo{author}{\bibfnamefont{H.~J.} \bibnamefont{Jensen}},
  \bibinfo{author}{\bibfnamefont{S.}~\bibnamefont{Lise}},
  \bibinfo{author}{\bibfnamefont{J.~M.} \bibnamefont{López}},
  \bibinfo{author}{\bibfnamefont{M.}~\bibnamefont{Nicodemi}},
  \bibinfo{author}{\bibfnamefont{J.-F.} \bibnamefont{Pinton}},
  \bibnamefont{and} \bibinfo{author}{\bibfnamefont{M.}~\bibnamefont{Sellitto}},
  \bibinfo{journal}{Phys. Rev. Lett.} \textbf{\bibinfo{volume}{84}},
  \bibinfo{pages}{3744} (\bibinfo{year}{2000}).

\bibitem[{\citenamefont{Pumir}(1996)}]{pumir_96}
\bibinfo{author}{\bibfnamefont{A.}~\bibnamefont{Pumir}},
  \bibinfo{journal}{Phys. Fluids} \textbf{\bibinfo{volume}{8}},
  \bibinfo{pages}{3112} (\bibinfo{year}{1996}).

\bibitem[{\citenamefont{Marino et~al.}(2012)\citenamefont{Marino,
  Sorriso-Valvo, D{'}{A}micis, Carbone, Bruno, and Veltri}}]{marino_12}
\bibinfo{author}{\bibfnamefont{R.}~\bibnamefont{Marino}},
  \bibinfo{author}{\bibfnamefont{L.}~\bibnamefont{Sorriso-Valvo}},
  \bibinfo{author}{\bibfnamefont{R.}~\bibnamefont{D{'}{A}micis}},
  \bibinfo{author}{\bibfnamefont{V.}~\bibnamefont{Carbone}},
  \bibinfo{author}{\bibfnamefont{R.}~\bibnamefont{Bruno}}, \bibnamefont{and}
  \bibinfo{author}{\bibfnamefont{P.}~\bibnamefont{Veltri}},
  \bibinfo{journal}{Astrophys. J.} \textbf{\bibinfo{volume}{750}},
  \bibinfo{pages}{41} (\bibinfo{year}{2012}).

\bibitem[{\citenamefont{Capet et~al.}(2008)\citenamefont{Capet,
 Mc{W}illiams, Molemaker, and Shchepetkin}}]{capet_08}
\bibinfo{author}{\bibfnamefont{X.}~\bibnamefont{Capet}},
  \bibinfo{author}{\bibfnamefont{J.~C.}~\bibnamefont{Mc{W}illiams}},
  \bibinfo{author}{\bibfnamefont{M.~J.}~\bibnamefont{Molemaker}}, \bibnamefont{and}
  \bibinfo{author}{\bibfnamefont{A.~F.}~\bibnamefont{Shchepetkin}},
  \bibinfo{journal}{J. Phys. Oceanogr.} \textbf{\bibinfo{volume}{38}},
  \bibinfo{pages}{29} (\bibinfo{year}{2008}).

\bibitem[{\citenamefont{Paoletti et~al.}(2008)\citenamefont{Paoletti, Fisher,
  Sreenivasan, and Lathrop}}]{paoletti_08}
\bibinfo{author}{\bibfnamefont{M.~S.} \bibnamefont{Paoletti}},
  \bibinfo{author}{\bibfnamefont{M.~E.} \bibnamefont{Fisher}},
  \bibinfo{author}{\bibfnamefont{K.~R.} \bibnamefont{Sreenivasan}},
  \bibnamefont{and} \bibinfo{author}{\bibfnamefont{D.~P.}
  \bibnamefont{Lathrop}}, \bibinfo{journal}{Phys. Rev. Lett.}
  \textbf{\bibinfo{volume}{101}}, \bibinfo{pages}{154501}
  (\bibinfo{year}{2008}).

\bibitem[{\citenamefont{Baggaley and Barenghi}(2011)}]{baggaley_11}
\bibinfo{author}{\bibfnamefont{A.~W.} \bibnamefont{Baggaley}} \bibnamefont{and}
  \bibinfo{author}{\bibfnamefont{C.~F.} \bibnamefont{Barenghi}},
  \bibinfo{journal}{Phys. Rev. E} \textbf{\bibinfo{volume}{84}},
  \bibinfo{pages}{067301} (\bibinfo{year}{2011}).

\bibitem[{\citenamefont{Sullivan and Patton}(2011)}]{sullivan_11}
\bibinfo{author}{\bibfnamefont{P.}~\bibnamefont{Sullivan}} \bibnamefont{and}
  \bibinfo{author}{\bibfnamefont{N.}~\bibnamefont{Patton}},
  \bibinfo{journal}{J. Atmos. Sci.} \textbf{\bibinfo{volume}{68}},
  \bibinfo{pages}{2395} (\bibinfo{year}{2011}).

\bibitem[{\citenamefont{Sun et~al.}(2012)\citenamefont{Sun, Mahrt, Banta, and
  Pichugina}}]{jielun_12}
\bibinfo{author}{\bibfnamefont{J.}~\bibnamefont{Sun}},
  \bibinfo{author}{\bibfnamefont{L.}~\bibnamefont{Mahrt}},
  \bibinfo{author}{\bibfnamefont{R.}~\bibnamefont{Banta}}, \bibnamefont{and}
  \bibinfo{author}{\bibfnamefont{Y.}~\bibnamefont{Pichugina}},
  \bibinfo{journal}{J. Atm. Sci.} \textbf{\bibinfo{volume}{69}},
  \bibinfo{pages}{338} (\bibinfo{year}{2012}).

\bibitem[{\citenamefont{Finnigan}(1999)}]{finnigan_99}
\bibinfo{author}{\bibfnamefont{J.}~\bibnamefont{Finnigan}},
  \bibinfo{journal}{Bound. Lay. Met.} \textbf{\bibinfo{volume}{90}},
  \bibinfo{pages}{529} (\bibinfo{year}{1999}).

\bibitem[{\citenamefont{de~Wiel et~al.}(2012)\citenamefont{de~Wiel, Moene, and
  Jonker}}]{jonker_12a}
\bibinfo{author}{\bibfnamefont{B.~J. H.~V.} \bibnamefont{de~Wiel}},
  \bibinfo{author}{\bibfnamefont{A.}~\bibnamefont{Moene}}, \bibnamefont{and}
  \bibinfo{author}{\bibfnamefont{H.}~\bibnamefont{Jonker}},
  \bibinfo{journal}{J. Atmos. Sci.} \textbf{\bibinfo{volume}{69}},
  \bibinfo{pages}{3097} (\bibinfo{year}{2012}).

\bibitem[{\citenamefont{Dr\"ue and Heinemann}(2007)}]{drue_07}
\bibinfo{author}{\bibfnamefont{C.}~\bibnamefont{Dr\"ue}} \bibnamefont{and}
  \bibinfo{author}{\bibfnamefont{G.}~\bibnamefont{Heinemann}},
  \bibinfo{journal}{Bound. Lay. Met.} \textbf{\bibinfo{volume}{124}},
  \bibinfo{pages}{361} (\bibinfo{year}{2007}).

\bibitem[{\citenamefont{Stevens and Feingold}(2009)}]{stevens_09}
\bibinfo{author}{\bibfnamefont{B.}~\bibnamefont{Stevens}} \bibnamefont{and}
  \bibinfo{author}{\bibfnamefont{G.}~\bibnamefont{Feingold}},
  \bibinfo{journal}{Nature} \textbf{\bibinfo{volume}{461}},
  \bibinfo{pages}{607} (\bibinfo{year}{2009}).

\bibitem[{\citenamefont{Vieillefosse}(1982)}]{vieille}
\bibinfo{author}{\bibfnamefont{P.}~\bibnamefont{Vieillefosse}},
  \bibinfo{journal}{J. Physique} \textbf{\bibinfo{volume}{43}},
  \bibinfo{pages}{837} (\bibinfo{year}{1982}).

\bibitem[{\citenamefont{Meneveau}(2011)}]{meneveau_11}
\bibinfo{author}{\bibfnamefont{C.}~\bibnamefont{Meneveau}},
  \bibinfo{journal}{Ann. Rev. Fluid Mech.} \textbf{\bibinfo{volume}{43}},
  \bibinfo{pages}{219} (\bibinfo{year}{2011}).

\bibitem[{\citenamefont{Polzin and Lvov}(2011)}]{polzin}
\bibinfo{author}{\bibfnamefont{K.}~\bibnamefont{Polzin}} \bibnamefont{and}
  \bibinfo{author}{\bibfnamefont{Y.}~\bibnamefont{Lvov}},
  \bibinfo{journal}{Rev. Geophys.} \textbf{\bibinfo{volume}{49}},
  \bibinfo{pages}{RG4003} (\bibinfo{year}{2011}).

\bibitem[{\citenamefont{Li and Meneveau}(2006)}]{li_meneveau_06}
\bibinfo{author}{\bibfnamefont{Y.}~\bibnamefont{Li}} \bibnamefont{and}
  \bibinfo{author}{\bibfnamefont{C.}~\bibnamefont{Meneveau}},
  \bibinfo{journal}{J. Fluid Mech.} \textbf{\bibinfo{volume}{558}},
  \bibinfo{pages}{133} (\bibinfo{year}{2006}).

\bibitem[{\citenamefont{Li}(2010)}]{li_10}
\bibinfo{author}{\bibfnamefont{Y.}~\bibnamefont{Li}}, \bibinfo{journal}{Physica
  D} \textbf{\bibinfo{volume}{239}}, \bibinfo{pages}{1948}
  (\bibinfo{year}{2010}).

\bibitem[{\citenamefont{Mininni et~al.}(2011)\citenamefont{Mininni, Rosenberg,
  Reddy, and Pouquet}}]{hybrid2011}
\bibinfo{author}{\bibfnamefont{P.}~\bibnamefont{Mininni}},
  \bibinfo{author}{\bibfnamefont{D.}~\bibnamefont{Rosenberg}},
  \bibinfo{author}{\bibfnamefont{R.}~\bibnamefont{Reddy}}, \bibnamefont{and}
  \bibinfo{author}{\bibfnamefont{A.}~\bibnamefont{Pouquet}},
  \bibinfo{journal}{Parallel Computing} \textbf{\bibinfo{volume}{37}},
  \bibinfo{pages}{316} (\bibinfo{year}{2011}).

\bibitem[{\citenamefont{Brethouwer et~al.}(2007)}]{brethouwer_07}
\bibinfo{author}{\bibfnamefont{G.}~\bibnamefont{Brethouwer}},
  \bibinfo{author}{\bibfnamefont{P.}~\bibnamefont{Billant}},
  \bibinfo{author}{\bibfnamefont{E.}~\bibnamefont{Lindborg}}, \bibnamefont{and}
  \bibinfo{author}{\bibfnamefont{J.-M.} \bibnamefont{Chomaz}},
  \bibinfo{journal}{J. Fluid Mech.} \textbf{\bibinfo{volume}{585}},
  \bibinfo{pages}{343} (\bibinfo{year}{2007}).

\bibitem[{\citenamefont{Billant and Chomaz}(2001)}]{billant_01}
\bibinfo{author}{\bibfnamefont{P.}~\bibnamefont{Billant}} \bibnamefont{and}
  \bibinfo{author}{\bibfnamefont{J.-M.} \bibnamefont{Chomaz}},
  \bibinfo{journal}{Phys. Fluids} \textbf{\bibinfo{volume}{13}},
  \bibinfo{pages}{1645} (\bibinfo{year}{2001}).

\bibitem[{\citenamefont{Lindborg}(2006)}]{lindborg2006}
\bibinfo{author}{\bibfnamefont{E.}~\bibnamefont{Lindborg}},
  \bibinfo{journal}{J. Fluid Mech.} \textbf{\bibinfo{volume}{550}},
  \bibinfo{pages}{207} (\bibinfo{year}{2006}).

\bibitem[{\citenamefont{Falkovich and Fouxon}(2005)}]{falko_05b}
\bibinfo{author}{\bibfnamefont{G.}~\bibnamefont{Falkovich}} \bibnamefont{and}
  \bibinfo{author}{\bibfnamefont{A.}~\bibnamefont{Fouxon}},
  \bibinfo{journal}{Phys. Rev. Lett.} \textbf{\bibinfo{volume}{94}},
  \bibinfo{pages}{214502} (\bibinfo{year}{2005}).

\bibitem[{\citenamefont{Lvov and Nazarenko}(2004)}]{lvov_04}
\bibinfo{author}{\bibfnamefont{Y.~V.} \bibnamefont{Lvov}} \bibnamefont{and}
  \bibinfo{author}{\bibfnamefont{S.}~\bibnamefont{Nazarenko}},
  \bibinfo{journal}{Phys. Rev. E} \textbf{\bibinfo{volume}{69}},
  \bibinfo{pages}{066608} (\bibinfo{year}{2004}).

\bibitem[{\citenamefont{Portelli et~al.}(2003)\citenamefont{Portelli,
  Holdsworth, and Pinton}}]{portelli_03}
\bibinfo{author}{\bibfnamefont{B.}~\bibnamefont{Portelli}},
  \bibinfo{author}{\bibfnamefont{P.~C.~W.} \bibnamefont{Holdsworth}},
  \bibnamefont{and} \bibinfo{author}{\bibfnamefont{J.-F.}
  \bibnamefont{Pinton}}, \bibinfo{journal}{Phys. Rev. Lett.}
  \textbf{\bibinfo{volume}{90}}, \bibinfo{pages}{104501}
  (\bibinfo{year}{2003}).

\bibitem[{\citenamefont{Fritts et~al.}(2009)\citenamefont{Fritts, Wang, and
  Werne}}]{fritts_09a}
\bibinfo{author}{\bibfnamefont{D.~C.} \bibnamefont{Fritts}},
  \bibinfo{author}{\bibfnamefont{L.}~\bibnamefont{Wang}}, \bibnamefont{and}
  \bibinfo{author}{\bibfnamefont{J.}~\bibnamefont{Werne}},
  \bibinfo{journal}{Geophys. Res. Lett.} \textbf{\bibinfo{volume}{36}},
  \bibinfo{pages}{396} (\bibinfo{year}{2009}).

\bibitem[{\citenamefont{Palmer}(2012)}]{palmer_12}
\bibinfo{author}{\bibfnamefont{T.~N.} \bibnamefont{Palmer}},
  \bibinfo{journal}{Q. J. R. Meteorol. Soc.} \textbf{\bibinfo{volume}{138}},
  \bibinfo{pages}{841} (\bibinfo{year}{2012}).

\bibitem[{\citenamefont{Gryanik et~al.}(2005)\citenamefont{Gryanik, Hartmann,
  Raasch, and Schr\"oter}}]{gryanik_05}
\bibinfo{author}{\bibfnamefont{V.~M.} \bibnamefont{Gryanik}},
  \bibinfo{author}{\bibfnamefont{J.}~\bibnamefont{Hartmann}},
  \bibinfo{author}{\bibfnamefont{S.}~\bibnamefont{Raasch}}, \bibnamefont{and}
  \bibinfo{author}{\bibfnamefont{M.}~\bibnamefont{Schr\"oter}},
  \bibinfo{journal}{J. Atmos. Sci.} \textbf{\bibinfo{volume}{62}},
  \bibinfo{pages}{2632} (\bibinfo{year}{2005}).

\end{thebibliography}
\end{document}